\documentclass[showpacs,twocolumn,prl,aps,color]{revtex4}
\usepackage{graphicx}
\begin{document}

\title{Midgap States in Antiferromagnetic Heisenberg Chains with A Staggered Field}

\author{Jizhong Lou and Changfeng Chen}
\affiliation{Department of Physics, University of Nevada, Las Vegas, Nevada
89154, USA}
\author{Jize Zhao, Xiaoqun Wang, Tao Xiang, Zhaobin Su, and Lu Yu}
\affiliation{Institute of Theoretical Physics, CAS, Beijing 100080, China}
\affiliation{The Interdisciplinary Center of Theoretical Studies, CAS,
Beijing 100080, China}

\date{\today}

\begin{abstract}
We study low-energy excitations in antiferromagnetic Heisenberg
chains with a staggered field which splits the spectrum into a
longitudinal and a transverse branch. Bound states are found to
exist inside the field induced gap in both branches.  They
originate from the  edge effects and are inherent to spin-chain
materials.  The sine-Gordon scaling $h_s^{2/3}|\log h_s|^{1/6}$
($h_s$: the staggered field) provides an accurate description for
the gap and midgap energies in the transverse branch for $S=1/2$
and the midgap energies in both branches for $S=3/2$ over a wide
range of magnetic field; however, it can  fit other low-energy
excitations only at much lower field.  Moreover, the integer-spin
$S=1$ chain displays scaling behavior that does not fit this
scaling law. These results reveal intriguing features of magnetic
excitations in spin-chain materials that deserve further
investigation.
\end{abstract}


\maketitle

Quantum spin chains have been a source of fascinating developments
in physics and many related fields since the early days of quantum
theory \cite{Beth,Faddeev,Hald,Osh1}. They  serve as model systems
for revealing fundamental physics in many materials. Progress in
experimental techniques has led to the synthesis and
characterization of many quasi-one-dimensional materials in the
last two decades. A very active field of theoretical and
experimental research has emerged for the study of novel magnetic
properties in low-dimensional spin systems where quantum
fluctuations play a crucial role \cite{Sachedev,Broholm}.

Intriguing low-energy phenomena like the spin gap induced by
magnetic field have been observed in many recently synthesized
materials such as Cu(C$_6$D$_5$COO)$_2$3D$_2$O \cite{Den2,Den1},
Yb$_4$As$_3$ \cite{Kohgi,Fulde,Osh2} and
CuCl$_2$¡¤2(dimethylsulfoxide)\cite{Ken04}.  The
Dzyaloshinskii-Moriya (DM) interaction is intrinsic to these
materials due to an alternating structure of molecular or
spin-orbital interactions \cite{Dyz,Mor}. Although it is one order
of magnitude smaller than the standard Heisenberg exchange
interaction in these materials, significant effects of the DM
interaction have been unveiled recently when the materials are
subjected to external magnetic fields \cite{Den2,Osh1,Zhao2003}.
An energy gap proportional to some power of the magnetic field
opens up.  These systems can be generally modeled by the following
Hamiltonian,
\begin{eqnarray}
\hat{H} = \sum_i \left( J\hat{\bf S}_i\cdot \hat{{\bf
S}}_{i+1}-(-)^i{\bf D} \cdot \hat{\bf S}_i\times \hat{\bf
S}_{i+1}\right. \nonumber \\
 \left. -\mu_B{\bf H}\cdot \left[{\bf g}^u+(-)^i{\bf
g}^s\right]\cdot \hat{\bf S}_i \right),
 \label{HSDMZ}
\end{eqnarray}
where the three terms in the summation are the antiferromagnetic
Heisenberg, DM and Zeeman splitting interactions, respectively.
While the exchange coupling constant $J$ and the uniform
(staggered) ${\bf g}^u$ (${\bf g}^s$) can be determined from the
neutron scattering and electron spin resonance experiments, the
${\bf D}$-vector can be obtained only through theoretical analysis
of the gap values in comparison with experiment
\cite{Osh1,Essler,Zhao2003}.

The DM term can be eliminated by performing a spin rotation about
the ${\bf D}$ vector by an angle $\alpha=\pm \frac 1 2$ $\tan^{-1}
(D/J)$ on the alternating sites. After neglecting anisotropic
terms in the limit of $D \ll J$, which are only needed to account
for the field dependence of the gap in different crystallographic
directions \cite{Zhao2003}, the relevant physics is captured by
the effective Hamiltonian \cite{Osh1},
\begin{equation}
\hat{H} =J \sum_i [\hat{\bf S}_i\cdot \hat{{\bf S}}_{i+1}-h_u \hat
S_i^x-(-)^ih_s\hat S_i^z], \label{effH}
\end{equation}
where  $h_u$ and $h_s$ are the effective dimensionless (scaled by
$g\mu_B/J$, and $J$ is taken as the energy unit hereafter)
uniform and staggered field, respectively. This
Hamiltonian has been mapped onto the quantum sine-Gordon model
using the bosonization technique \cite{Osh1} and the
aforementioned novel magnetic properties are described in terms of
soliton, antisoliton, and breather excitations
\cite{EsslerTsvelik98,Osh1,Essler,Ken04}.

Despite intensive investigations, quasi-one-dimensional spin
systems with  DM interaction continue to display surprising new
physics. In this paper we show that the excitation spectra are
split into a longitudinal and a transverse branch, while edge
effects result in midgap states in both branches for open chains
with both integer and half-integer spins in the thermodynamic
limit. These midgap states are inherent to real spin-chain
materials which always have chain ends.  The existing field theory
is established in the context of the quantum sine-Gordon model for
periodic spin-1/2 chains \cite{Osh1,EsslerTsvelik98}. It is very
important to clarify how soliton, antisoliton, and breather modes
in the field theory are related to the gap and midgap states in
different branches. Recent electron spin resonance experiment
\cite{Zvy04} on a spin-1/2 chain compound
[Pyrimidine-Cu(NO$_3$)$_2$(H$_2$O )$_2$]$_n$(CuPM) revealed that
some excitation modes, including one of the most intensive modes,
cannot be fully described in terms of the quantum sine-Gordon
model \cite{Osh1}. This raises fundamental issues regarding the
nature of  these magnetic excitations. Moreover, current
discussions focus almost exclusively on the $S=1/2$ case; it is of
great interest to examine the low-energy excitations of
higher-spin chains \cite{Zz,Lou} and their possible connections to
the spin-1/2 chain and the field theoretical description.

We use the density matrix renormalization group (DMRG) method
\cite{white,peschel} to study directly the Hamiltonian defined by
Eq. (\ref{effH}) for both integer ($S$=1) and half-integer
($S$=1/2 and 3/2) spin chains.  We find the same qualitative behavior
for the gap and midgap states over a wide range of $h_u$ and,
consequently, focus below on the effect of $h_s$ at $h_u$=0.
Unless explicitly stated otherwise, we use open boundaries in all
calculations. The results are obtained by keeping 400 states, but
examined at low fields with 800 states.  The truncation errors are
on the order of $10^{-7}$ for $S=3/2$ and much smaller for $S$=1/2
and 1.  The gap and midgap energies in the thermodynamic limit were
obtained by extrapolating numerical results for up to 200 sites to
the long-chain limit.

\begin{figure}[h!]
\includegraphics[width=6.5cm,angle=0]{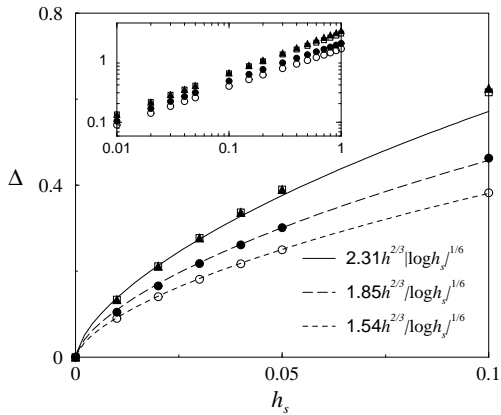}
\caption{The staggered field dependence of the transverse and
longitudinal gaps ($\Delta_T$: solid circles; $\Delta_L$: solid
triangles) and the midgap-state energies ($\Delta^m_T$: open
circles; $\Delta^m_L$: open squares) for the spin-1/2 chain.  The
lines are fittings to the $h_s^{2/3}|\log h_s|^{1/6}$ scaling
behavior anticipated from the conformal field theory \cite{Osh1}.
The slopes of the plots which determine, up to log corrections,
the scaling exponents of the leading order term are shown in the
inset with the same axis labels, in the logarithmic scale.  Data
show a common scaling exponent of $\gamma=2/3$ when the field is
not too high.} \label{fig1_mgs}
\end{figure}

Figure \ref{fig1_mgs} shows the low-lying energies relative to the
non-degenerate ground state for the spin-1/2 chain.  At $h_u=0$
the spectrum splits into a transverse branch with $|S_{tot}^z|$=1
and a longitudinal branch with $S_{tot}^z$=0, with gaps $\Delta_T$
and $\Delta_L$, respectively, determined by scaling analyses that
confirm the continuous nature of each spectrum in the large-$L$
limit.  Midgap states appear in both branches.  The gaps and the
midgap-state energies $\Delta^m_T$ and $\Delta^m_L$ are obtained
from the extrapolations of the DMRG data as shown in Fig. 2.

\begin{figure}[h!]
\includegraphics[width=6.3cm,angle=0]{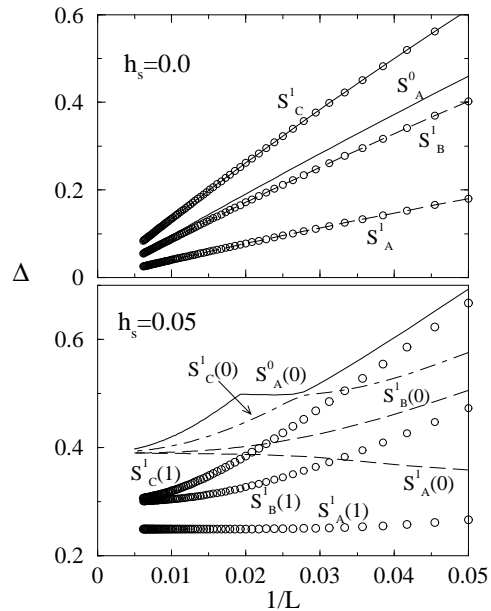}
\caption{The excitation energies versus $1/L$ for several lowest states
of the spin-1/2 chain labeled by $S^{S_{tot}}_{\alpha}$ (upper panel)
and $S^{S_{tot}}_{\alpha}(|S^z_{tot}|)$ (lower panel) [$\alpha=A, B, C$].
They all scale to zero as $L\rightarrow\infty$ at $h_s=0$.  At finite
$h_s$ ($h_s=0.05$ shown), they split into a longitudinal (lines) and a
transverse (symbols) branch with distinct gap and midgap states
($S_A^1(1)\rightarrow \Delta_T^m$, $S_B^1(1), S_C^1(1)\rightarrow\Delta_T$,
$S_A^1(0), S_B^1(0)\rightarrow\Delta_L^m$, $S_C^1(0)\rightarrow \Delta_L$,
see the text for details).}
\label{fig2_mgs}
\end{figure}

The conformal field theory for the spin-1/2 chain under a
staggered magnetic field predicted \cite{Osh1} the scaling
behavior $h_s^{2/3}|\log h_s|^{1/6}$ for the induced gap.  We
apply this scaling form to our numerical results as shown in Fig.
1.  It is interesting to note that the field theory not only
provides a good description in the low-field region for the
transverse gap, a periodic chain feature, as expected, but also
fits well the transverse midgap, a genuine open boundary edge
effect (see below).  The scaling exponents at the low field limit
share the common value of 2/3 predicted by the field theory.
However, the fitting range is different for the two branches: the
deviation from the predicted scaling behavior starts at
$h_s\approx0.03$ in the longitudinal branch while a satisfactory
fitting persists beyond $h_s$=0.1 for the transverse branch.
Moreover, the midgap state appears in the transverse branch as
soon as $h_s$ is applied, but it does not become distinguishable
in the longitudinal branch until $h_s=0.1$ which corresponds to a
field of $14$ Tesla in copper benzoate \cite{Den2}.  It suggests
that the midgap state in the transverse branch should be
observable at low fields, while much higher fields are required
for its identification in the longitudinal branch. Since the
midgap states are due to the chain-end effect, their intensity is
proportional to the impurity/defect concentration. Samples with
high level of nonmagnatic impurity doping may be needed for their
detection.

To examine the origin of the midgap states, we consider a bond
impurity model \cite{wang1996}.  A coupling $J'$ is added between
the two end spins and varied from 0 (open boundary) to $J$
(periodic boundary). The midgap states develop in each branch when
$J'$ deviates from $J$, indicating that these states appear as a
consequence of the edge effect.  The same effect is also observed
in all higher-spin chains discussed below.  Alternatively, we show
in Fig. 2 for the spin-1/2 case that the midgap states can be
traced back to the three lowest  excitation states with ${\bf
S}_{tot}=1$ at $h_s$=0, labeled as $S^1_A$, $S^1_B$ and $S^1_C$,
respectively. A comparison of the $h_s=0$ and 0.05 data shows that
state $S^1_A$ splits into  midgap states $S^1_A(1)$ and
$S^1_A(0)$, state $S^1_B$ splits into $S^1_B(1)$ that becomes the
bottom of the transverse continuum and $S^1_B(0)$ that is a midgap
state degenerate with $S^1_A(0)$, and state $S^1_C$ splits into
$S^1_C(1)$  degenerate with $S^1_B(1)$ and $S^1_C(0)$ which is the
bottom of the longitudinal continuum. The lowest excitation state
$S^0_A$ with ${\bf S}_{tot}=0$ moves higher in energy and enters
the longitudinal continuum at $h_s>0$, which mixes  two states
$S^0_A(0)$ and $S^1_C(0)$ at about $L=36$.

\begin{figure}[h!]
\includegraphics[width=6.0cm,angle=0]{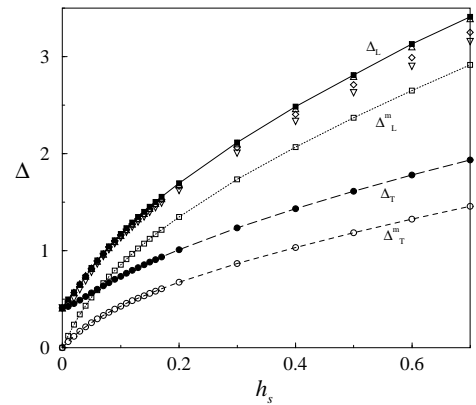}
\caption{The staggered field dependence of the transverse and
longitudinal gaps ($\Delta_T$: solid circles; $\Delta_L$: solid
squares) and the midgaps ($\Delta^m_T$: open circles,
$|S_{tot}^z|=1$; $\Delta^m_L$: open squares, $S_{tot}^z=0$;
down-triangles: $S_{tot}^z=0$, diamonds: $S_{tot}^z=0$, and
up-triangles: $|S_{tot}^z|=2$) in the spin-1 chain.  The lines are
guide to the eyes only. The midgap states falling off from the
longitudinal continuum are not connected by lines for clarity.}
\label{fig3_mgs}
\end{figure}

We now turn to the spin-1 case where the spectrum is gapful at
$h_s=0$. In the thermodynamic limit, the ground state of the open
chain is four-fold degenerate owing to the topological edge effect
as interpreted in the valence-bond-solid (VBS) picture
\cite{VBS1}. The staggered field splits the four-fold degenerate
ground state into three mixed states: a $ S_{tot}^z=0$ ground
state, a doubly degenerate $|S_{tot}^z|=1$ transverse midgap state
and a $S_{tot}^z=0$ longitudinal state. The dependence of the gap
and midgap-state energies on $h_s$ is shown in Fig.
\ref{fig3_mgs}. In addition to the midgap states originating from
a mixture of the $|S_{tot}^z|=0, ~1$ states in the ground-state
manifold, there are also midgap states with $S_{tot}^z=0$ and
$|S_{tot}^z|=2$ below but close to the bottom of the longitudinal
continuum.  We have tried fitting the  results to the field
theoretical formula, but found that the $h_s^{2/3}|\log
h_s|^{1/6}$ scaling is unsuitable for $S=1$ in any field range.
While this is not surprising for the gaps at low field since they
have finite values at $h_s=0$, the fact that it does not apply to
the midgap energies that do scale to zero at $h_s=0$ strongly
suggests different intrinsic behavior for integer and half-integer
spin cases \cite{Hald}.

\begin{figure}[h!]
\includegraphics[width=6.5cm,angle=0]{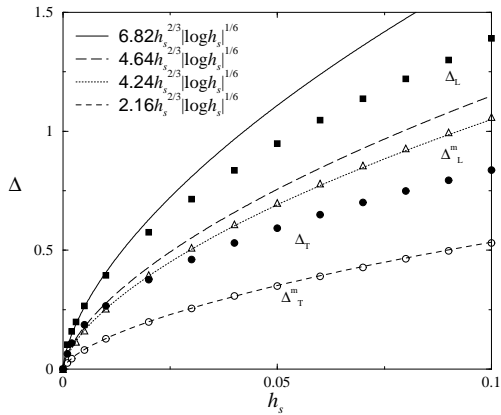}
\caption{The staggered field dependence of the transverse and
longitudinal gaps ($\Delta_T$: solid circles; $\Delta_L$: solid
squares) and the corresponding lower midgap-state energies
($\Delta_T^m$: open circles; $\Delta_L^m$: open triangles) in the
spin-3/2 chain.  The lines are  fittings to $h_s^{2/3}|\log
h_s|^{1/6}$ for these low-energy excitations.} \label{fig4_mgs}
\end{figure}

Figure \ref{fig4_mgs} shows the scaling behavior of the gap and
midgap-state energies of the spin-3/2 chain.  The low-field
sine-Gordon scaling derived for the spin-1/2 chain also fits these
results.  In particular, the two lowest midgap states are well
described beyond $h_s=0.1$.  It highlights common central charge
\cite{karen} and topological features \cite{to1} intrinsic to all
half-integer-spin chains.  However,  subtle yet important
differences exist between the $S=1/2$ and $S=3/2$ cases.  Unlike
the spin-1/2 case where bound midgap state develops at low field
explicitly in the transverse branch only, midgap states appear in
both branches at very low fields for $S=3/2$.  Moreover, in the
spin-3/2 case the $h_s^{2/3}|\log h_s|^{1/6}$ scaling fits both
gaps only in the low field limit ($h_s<0.02$), while the fitting
works very well for the two lower midgap states in both branches
over a much larger field range.  Meanwhile, similar to the $S$=1
case, additional midgap states fall off from the bottom of the
longitudinal continuum (not shown in Fig. 4 for clarity since a
few of them are very close to the bottom at low fields). These
 similarities and distinctions between different
half-integer spin chains deserve further studies.

Finally, we remark on the nature of the midgap states that share
some common topological origin for all spin magnitudes.  Although
the gap at $h_s=0$ is considered a fundamental criterion to
distinguish the integer and half-integer spin cases
\cite{Beth,Hald}, it has been shown that an extra Berry phase
contribution reflecting the topological feature leads to edge
states in all open spin chains with $S > 1/2$ \cite{Ng}.  Most
recently, it was also shown \cite{to1} that a topological string
order originally derived \cite{to3,to4} for integer-spin chains
exists in half-integer-spin chains as well.  Nonetheless, the
topological edge effect has been experimentally identified so far
only in integer-spin chain materials where the excitation spectrum
is gapful and topological excitations unstable against external
disturbances (such as doping), can be measured \cite{wang1996}.
For the $S=1$ case, this effect was observed in
Y$_{1-x}$BaCa$_x$NiO$_5$ \cite{Ditusa} and
Y$_2$BaNi$_{1-x}$Mg$_x$O$_5$ \cite{Ken03}.  The present results
unveil the topological edge effects driven by the staggered
magnetic field.  The midgap states split off from the degenerate
ground state of $S=1$ chain (and all other integer-spin chains)
represent a secondary (field induced) topological edge effect
since bound midgap states are already observable at $h_s=0$.
Meanwhile, additional bound states fall inside the field induced
gap from the continuum spectrum of all spin chains with  $S\geq 1$
at $h_s>0$.

It should be noted that the staggered field $h_s$ arises from a
nonzero uniform field $h_u$ in real materials.  The latter does
not change the qualitative scaling behavior when the field is not
too strong \cite{Osh1,Zhao2003}.  We have examined the gap and
midgap states at nonzero $h_u$ and found the same qualitative
results at realistic $h_s$/$h_u$ ratios although the classification
of the longitudinal and transverse branches no longer holds strictly
because of the breaking of the axial symmetry by $h_u$.

In summary, we have shown that bound midgap states generally exist
in open boundary antiferromagnetic Heisenberg chains with a
staggered magnetic field. They should appear in
quasi-one-dimensional materials with the Dzyoloshinskii-Moriya
interaction.  Some of the gap and midgap energies for the
half-integer spin chains fit well to a scaling function derived
from the quantum sine-Gordon model.  However, it is revealed that
(i) other low-energy excitations of the half-integer spin chains
do not fit equally well and (ii) the scaling behavior of the
integer spin chain is qualitatively different. Further
experimental and theoretical investigations are called for to
fully identify low energy excitations, especially the midgap
states as a general phenomenon in open spin chains.

We thank M. Kenzelmann, A.K. Kolezhuk, A.A. Nersesyan, S.J. Qin
and Y.J. Wang for helpful comments. This work was supported by the
National Basic Research Program under the Grant 2005CB32170X and
the NSFC Under Grant No. 10425417 and 90203006, and by the U. S.
Department of Energy under Cooperative Agreement DE-FC52-01NV14049
at UNLV. CFC acknowledges the hospitality and support of ICTS.

\end{document}